\documentstyle[12pt,epsfig]{article}
\begin{document}

\begin{center}

{\LARGE\bf Afterglows from Jetted GRB Remnants}

\vspace{5mm}

{\large\bf HUANG Yong-feng}

\vspace{2mm}

{\small\bf Department of Astronomy, Nanjing University,
        Nanjing 210093, P. R. China; hyf@nju.edu.cn} 

{\small\bf Department of Physics, the University of Hong Kong,
                     Hong Kong, P. R. China}

\end{center}

\begin{abstract}

We have found that the conventional generic dynamical model for gamma-ray
bursts (GRBs) cannot reproduce the Sedov solution in the non-relativistic
limit. Based on our refined generic dynamical model, we investigate
afterglows from jetted GRB remnants numerically. Many new results are
reached. For example, we find no obvious break in the optical light
curve during the relativistic phase itself. But an obvious break does
exist at the transition from the relativistic phase to the non-relativistic
phase, which typically occurs at time 10 to 30 days. It is very interesting
that the break is affected by many parameters, such as the electron energy
fraction, $\xi_{\rm e}$, the magnetic energy fraction, $\xi_{\rm B}^2$, 
the initial opening angle of the jet,$\theta_0$, and the medium density, $n$. 
We also find that afterglows from jetted GRB
remnants are uniformly characterized by a quick decay during the
non-relativistic phase, with power law timing index greater than 2.1 .
Afterglows from GRB 970228, 980326, 980519, 990123, 990510 and 991208 can
be satisfactorily fitted if the corresponding GRB ejecta are highly
collimated.

\end{abstract}

\vspace{0.5cm}

\section{Introduction}

The discovery of afterglows from some gamma-ray bursts (GRBs) has
opened up a new era in the field (Costa et al. 1997; Galama et al. 1999). 
The so called fireball model (M\'{e}sz\'{a}ros \& Rees 1992; 
Sari et al. 1996), which can successfully explain the major features of
GRB afterglows (Vietri 1997; Waxman 1997; Tavani 1997; Wijers et al. 1997),
becomes the most popular model (see Piran 1999 and van Paradijs et 
al. 2000 for recent 
reviews). Some GRBs localized by BeppoSAX satellite
have implied isotropic energy release of more than $10^{54}$ ergs 
(Madsen et al. 1999; Harrison et al. 1999),
this has forced many theorists to deduce that GRB radiation must be
highly collimated (Groot et al. 1998; Owens et al. 1998;
Halpern et al. 1999;  Castro-Tirado et al. 1999; Sari et al. 1999;
Wei \& Lu 1999a, b).

\begin{figure}[htp]
\centerline{\epsfig{file=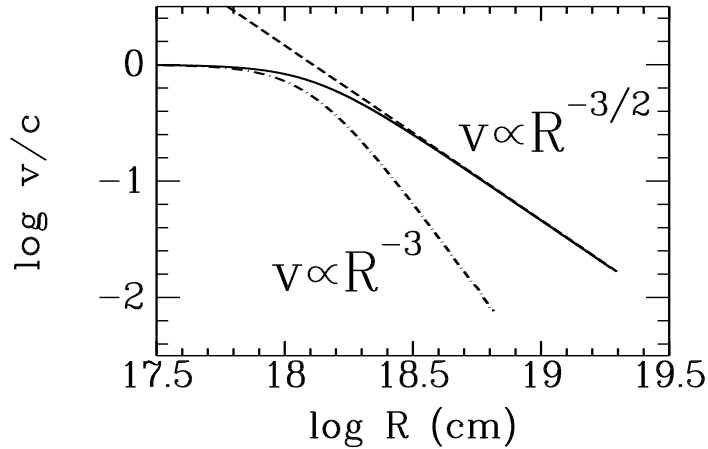, angle=-90, height=50.0mm, width=6.8cm, 
bbllx=150pt, bblly=165pt, bburx=510pt, bbury=570pt}}
\begin{center}
\begin{minipage}{12cm}
\caption[]{Velocity vs. radius for an isotropic
adiabatic fireball. The dashed line is the familiar Sedov solution in 
the Newtonian phase. The dash-dotted line is drawn according to Eq.~(1),
which differs from the dashed line markedly. The solid line
corresponds to our refined model (i.e., Eq.~(2)), which is
consistent with the Sedov solution.}
\label{ssmsbengps}
\end{minipage}
\end{center}
\end{figure}

To tell a jet from an isotropic fireball, it seems that we must
resort to the afterglow light curves. When the bulk Lorentz factor of
a jet drops to $\gamma < 1 / \theta$, with $\theta$ the half opening
angle, the edge of the jet becomes visible, the light curve will
steepen by $t^{-3/4}$. This is called the edge effect (M\'{e}sz\'{a}ros
\& Rees 1999; Panaitescu \& M\'{e}sz\'{a}ros 1999; Kulkarni et al. 1999). 
Additionally, the lateral expansion of a relativistic
jet will make the break even more precipitous (Rhoads 1999; Lamb 2000). 
So it is generally believed that afterglows from jetted GRBs are
characterized by an obvious break in the light curve at
{\em the relativistic stage}.

In this article, I use our refined dynamical model to study the jet
effect on the afterglow light curves.

\section{Our Refined Generic Model}

The importance of the non-relativistic phase of fireball expansion has
been stressed by Huang et al. (1998a, b; also see Wijers et al. 1997). 
In the literature, it is
generally believed that the following equation can depict the evolution
of GRB remnants (Chiang \& Dermer 1999)
\begin{equation}
\label{dgdm1}
\frac{d \gamma}{d m} = - \frac{\gamma^2 - 1}{M},
\end{equation}
where $m$ is the rest mass of the swept-up medium, $M$ is the total mass
in the co-moving frame, including internal energy. However, Huang
et al. (1999) pointed out that during the non-relativistic phase
of an adiabatic expansion, Eq.~(1) cannot reproduce the familiar Sedov
solution. This is clearly shown in Fig.~1.

  \begin{figure}[htb]
  \begin{center}
  \leavevmode
  \centerline{ \hbox{ \hbox{} \hspace{0.5in}
  \epsfig{figure=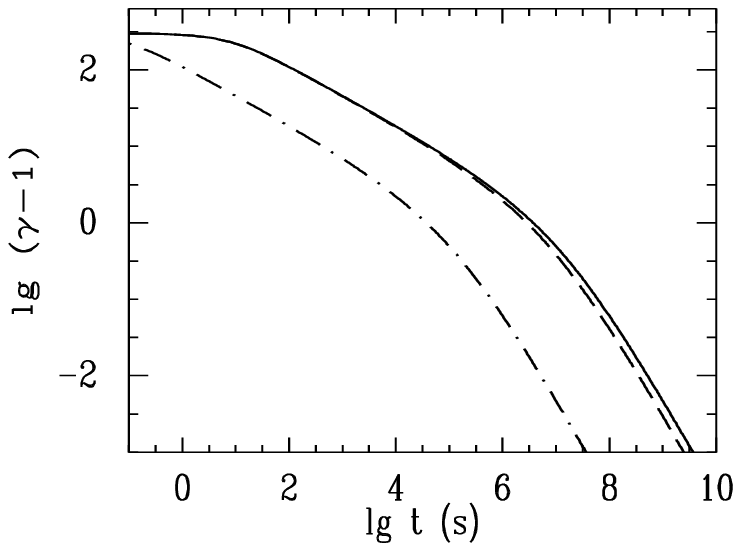,width=2.38in,height=1.2in,angle=-90,
bbllx=230pt, bblly=380pt, bburx=335pt, bbury=580pt}
  \hspace{0.5in}
  \epsfig{figure=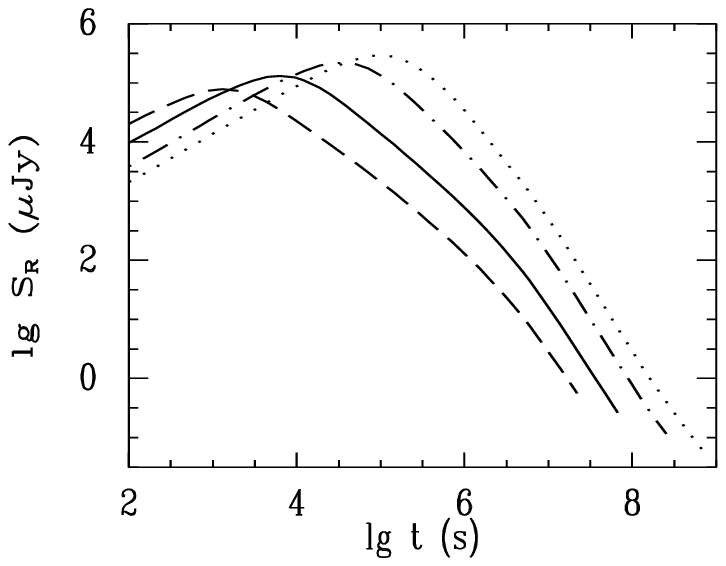,width=2.38in,height=1.2in,angle=-90,
bbllx=230pt, bblly=350pt, bburx=335pt, bbury=550pt}
  }}
  \begin{flushright}
  \parbox[t]{2.9in} { \caption { \label{ssrhor} Evolution 
of $\gamma$. The solid line
corresponds to a jet with ``standard'' parameters. Other lines are drawn
with only one parameter altered: the dashed line corresponds 
to $\theta_0 = 0.1$, and the dash-dotted line corresponds 
to $n = 10^6$ cm$^{-3}$ (Huang et al. 2000c). 
  }} \ \hspace{.2in} \
  \parbox[t]{2.9in} { \caption { \label{ssrhowd} The 
effect of $\xi_{\rm e}$ on the R-band light
curve. $S_{\rm R}$ is the R-band flux density. 
The solid line corresponds to a jet with ``standard'' parameters.
Other lines are drawn with only $\xi_{\rm e}$ altered. The dashed,
dash-dotted, and dotted lines correspond to $\xi_{\rm e} = 0.03$, 0.4, 
and 1.0 respectively (Huang et al. 2000c). }}
  \end{flushright}
  \end{center}
  \end{figure}

Huang et al. (1999) have proposed a refined equation
\begin{equation}
\label{dgdm2}
\frac{d \gamma}{d m} = - \frac{\gamma^2 - 1}
       {M_{\rm ej} + \epsilon m + 2 ( 1 - \epsilon) \gamma m}, 
\end{equation}
where $M_{\rm ej}$ is the initial baryon mass ejected from the GRB
central engine, and $\epsilon$ is the radiative efficiency. For an
adiabatic fireball, $\epsilon = 0$; and for a highly radiative one,
$\epsilon = 1$. Huang et al. (1999) have shown that Eq.~(2) is 
correct for both radiative and adiabatic fireballs, and in both
ultra-relativistic and non-relativistic phases (c.f. Fig.~1).

\section{Evolution of Jetted GRB Remnants}

Using our refined dynamical model, the evolution of the beamed ejecta
can be described by (Huang et al. 2000a, b, c, d):
\begin{equation}
\label{drdt1}
\frac{d R}{d t} = \beta c \gamma (\gamma + \sqrt{\gamma^2 - 1}),
\end{equation}
\begin{equation}
\label{dmdr2}
\frac{d m}{d R} = 2 \pi R^2 (1 - \cos \theta) n m_{\rm p},
\end{equation}
\begin{equation}
\label{dthdt3}
\frac{d \theta}{d t} = \frac{c_{\rm s} (\gamma + \sqrt{\gamma^2 - 1})}{R},
\end{equation}
\begin{equation}
\label{dgdm4}
\frac{d \gamma}{d m} = - \frac{\gamma^2 - 1}
       {M_{\rm ej} + \epsilon m + 2 ( 1 - \epsilon) \gamma m}, 
\end{equation}
\begin{equation}
\label{cs5}
c_{\rm s}^2 = \hat{\gamma} (\hat{\gamma} - 1) (\gamma - 1) 
	      \frac{1}{1 + \hat{\gamma}(\gamma - 1)} c^2 , 
\end{equation}
where $m_{\rm p}$ is the proton mass, $c_{\rm s}$ is the co-moving 
sound speed, $\hat{\gamma} \approx (4 \gamma + 1)/(3 \gamma)$ is the 
adiabatic index, and $c$ is the speed of light. Below we will  
consider only adiabatic jets, for which $\epsilon \equiv 0$).  

  \begin{figure}[htb]
  \begin{center}
  \leavevmode
  \centerline{ \hbox{ \hbox{} \hspace{0.5in}
  \epsfig{figure=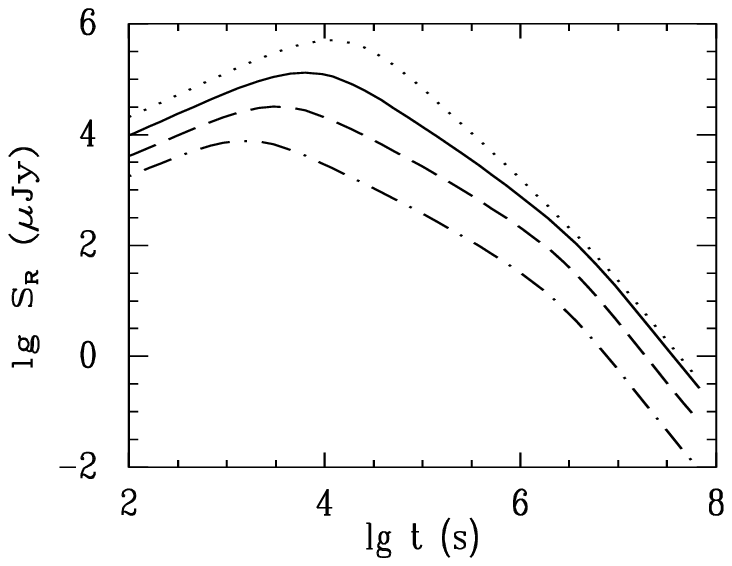,width=2.38in,height=1.2in,angle=-90,
bbllx=230pt, bblly=380pt, bburx=335pt, bbury=580pt}
  \hspace{0.5in}
  \epsfig{figure=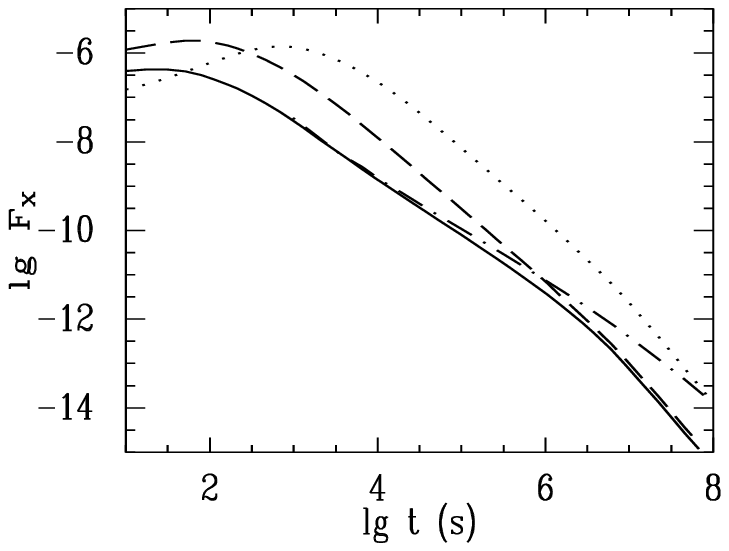,width=2.38in,height=1.2in,angle=-90,
bbllx=230pt, bblly=350pt, bburx=335pt, bbury=550pt}
  }}
  \begin{flushright}
  \parbox[t]{2.4in} { \caption { \label{ssrhor} The 
effect of $\xi_{\rm B}^2$ on the
optical light curve. The solid line corresponds to a jet with
``standard'' parameters. Other lines are drawn with only $\xi_{\rm B}^2$
altered. The dash-dotted, dashed, and the dotted lines are for
$\xi_{\rm B}^2 = 10^{-4}, 10^{-3}$, and $10^{-1}$ respectively 
(Huang et al. 2000c). }} 
\ \hspace{.2in} \
  \parbox[t]{3.4in} { \caption { \label{ssrhowd} The 
effect of $\xi_{\rm e}, \xi_{\rm B}^2$ and $\theta_0$ 
on the X-ray light curve. $F_{\rm X}$ is 0.1 --- 10 keV band flux,
in units of 
ergs$\cdot$cm$^{-2}\cdot$s$^{-1}$. The solid line corresponds to a jet 
with ``standard'' parameters. Other lines are drawn with only one 
parameter altered. The dash-dotted line is for $\theta_0 = 1.3$, 
the dashed line is for $\xi_{\rm B}^2 = 0.1$, and the dotted line is 
for $\xi_{\rm e} = 1$ (Huang et al. 2000d).}}
  \end{flushright}
  \end{center}
  \end{figure}

A strong blastwave will be generated due to the interaction of the 
jet and the ISM. Synchrotron radiation from the shock accelerated 
ISM electrons gives birth to afterglows. As usual we assume that 
the magnetic energy density in the co-moving frame is a 
fraction $\xi_{\rm B}^2$ of the total thermal energy density 
($B'^2 / 8 \pi = \xi_{\rm B}^2  e'$), and that electrons carry 
a fraction $\xi_{\rm e}$ of the proton energy. This means that the 
minimum Lorentz factor of the random motion of electrons in 
the co-moving frame is 
$\gamma_{\rm e,min} = \xi_{\rm e} (\gamma - 1) 
		     m_{\rm p} (p - 2) / [m_{\rm e} (p - 1)] + 1$, 
where $p$ is the index characterizing the power law energy distribution 
of electrons, and $m_{\rm e}$ is the electron mass. 

\section{Numerical Results}

For convenience, let us define the following initial values or parameters 
as a set of ``standard'' parameters: initial energy per solid 
angle $E_0 / \Omega_0 = 10^{54}$ ergs/$4 \pi$, $\gamma_0 = 300$,
$n = 1$ cm$^{-3}$, $\xi_{\rm B}^2 = 0.01$, $p = 2.5$, 
$\xi_{\rm e} =0.1$, $\theta_0 = 0.2, \theta_{\rm obs}=0$, 
$D_{\rm L} = 10^6$ kpc, where $\theta_{\rm obs}$ is 
the angle between the line of sight and the jet axis, and $D_{\rm L}$ 
is the luminosity distance. 

We have followed the evolution of jetted GRB remnants numerically and 
calculated their afterglows (Huang et al. 2000a, c, d).
Fig. 2 shows the evolution of $\gamma$
for some exemplary jets. We see that the ejecta will cease 
to be highly relativistic at time $t \sim 10^5$ --- $10^6$ s. 
This gives strong support to our previous argument that we should
be careful in discussing the fireball evolution under the simple 
assumption of ultra-relativistic limit (Huang et al. 1998a, b, 1999,
2000a, b, c, d).

Fig. 3 illustrates the effect of $\xi_{\rm e}$ on the optical 
(R-band) light curves. In no case could we observe the theoretically 
predicted light curve steepening (with the break point determined 
by $\gamma \sim 1/\theta$) during {\em the relativistic stage itself}.
The reason is: at
time of $\gamma \sim 1/\theta$, the jet is already in its mildly 
relativistic phase and it will become non-relativistic soon after that, 
so the break due to the edge effect and the lateral expansion
effect does not have time to emerge during the relativistic 
phase (Huang et al. 2000a, c).
Further more, since $\gamma$ is no longer much larger
than 1, conventional theoretical analyses (under the assumption 
of $\gamma \gg 1$) are not proper. However, when $\xi_{\rm e}$ is 
small, an obvious break does appear in the light curve, but it is clearly 
due to the relativistic-Newtonian transition.  The simulation by 
Moderski et al. (2000) does not reveal such breaks, because their model 
is not appropriate for Newtonian expansion. When $\xi_{\rm e}$ is 
large, the break disappears. This is not difficult to understand.
According to the analysis in the ultra-relativistic limit, the 
time that the light curve peaks scales as 
$t_{\rm m} \propto \xi_{\rm e}^{4/3} (\xi_{\rm B}^2)^{1/3}$.
Fig. 3 shows this trend qualitatively. 
In the case of $\xi_{\rm e} = 1.0$, $t_{\rm m}$ is as large 
as $\sim 10^5$ s, then we can not see the initial power law 
decay (with timing index $\alpha \sim 1.1$) in the relativistic 
phase, it is hidden by the peak.
So the break disappears. We should also note that in 
all cases, light curves during the non-relativistic phase 
are characterized by quick decays, with $\alpha \geq 2.1$. 
This is quite different from isotropic fireballs, whose light curves 
steepen only slightly after entering the Newtonian phase 
(i.e., $\alpha \sim 1.3$). 

Fig. 4 illustrates the effect of $\xi_{\rm B}^2$ on the optical 
light curves. Interestingly but not surprisingly, we see that 
$\xi_{\rm B}^2$  has an effect similar to $\xi_{\rm e}$: for small 
$\xi_{\rm B}^2$ values, there are obvious breaks at the 
relativistic-Newtonian transition points; but for 
large $\xi_{\rm B}^2$ values, the break disappears, 
we could only observe a single steep line with $\alpha \geq 2.1$. 
This is also due to the dependence of $t_{\rm m}$ on $\xi_{\rm B}^2$ 
(Huang et al. 2000c). 

We have also investigated the effects of other parameters such 
as $\theta_0, n, p$ on the optical light curves (Huang et al. 2000c). 
When $\theta_0 \geq 0.4$, the light curve becomes very similar to 
that of an isotropic fireball and no steep break exists. In the 
case of a dense medium ($n \geq 10^3$ cm$^{-3}$), the expansion 
becomes non-relativistic quickly, so that the break also disappears 
and we could only observe a quick decay with $\alpha \geq 2.1$. But 
generally speaking, $p$ does not affect the break notably. 

In Fig. 5, we show the effects of $\xi_{\rm e}, \xi_{\rm B}^2$ 
and $\theta_0$ on the X-ray afterglow light curves. Again we see 
that a large value of any of these parameters will make the 
relativistic-Newtonian break disappear. X-ray afterglows from 
GRB 980519 decayed very quickly, with $\alpha \approx 1.83 \pm 0.3$. 
We suggest that they may be produced by a jet with a large 
$\xi_{\rm B}^2$ or a large $\xi_{\rm e}$ (Huang et al. 2000d).

\section{Comparison with Observations}

Optical afterglows from GRB 990123, 990510 are characterized by
an obvious break in the light curve, and afterglows from
GRB 970228, 980326, 980519, 991208 faded rapidly. We suggest that
these phenomena are due to beaming effects. We have fitted these
afterglows based on our refined jet model and find that the
observations can be reproduced easily with a universal initial
half opening angle $\theta_0 \sim 0.1$ (Fig. 6). 
The obvious light curve
break in GRB 990123 is due to the relativistic-Newtonian transition
of the beamed ejecta, and the rapidly fading afterglows come from
synchrotron emissions during the mildly relativistic and
non-relativistic phases. We thus strongly suggest that the rapid
fading of afterglows currently observed in some GRBs is evidence
for beaming in these cases.

  \begin{figure}[htb]
  \begin{center}
  \leavevmode
  \centerline{ \hbox{ \hbox{} \hspace{0.5in}
  \epsfig{figure=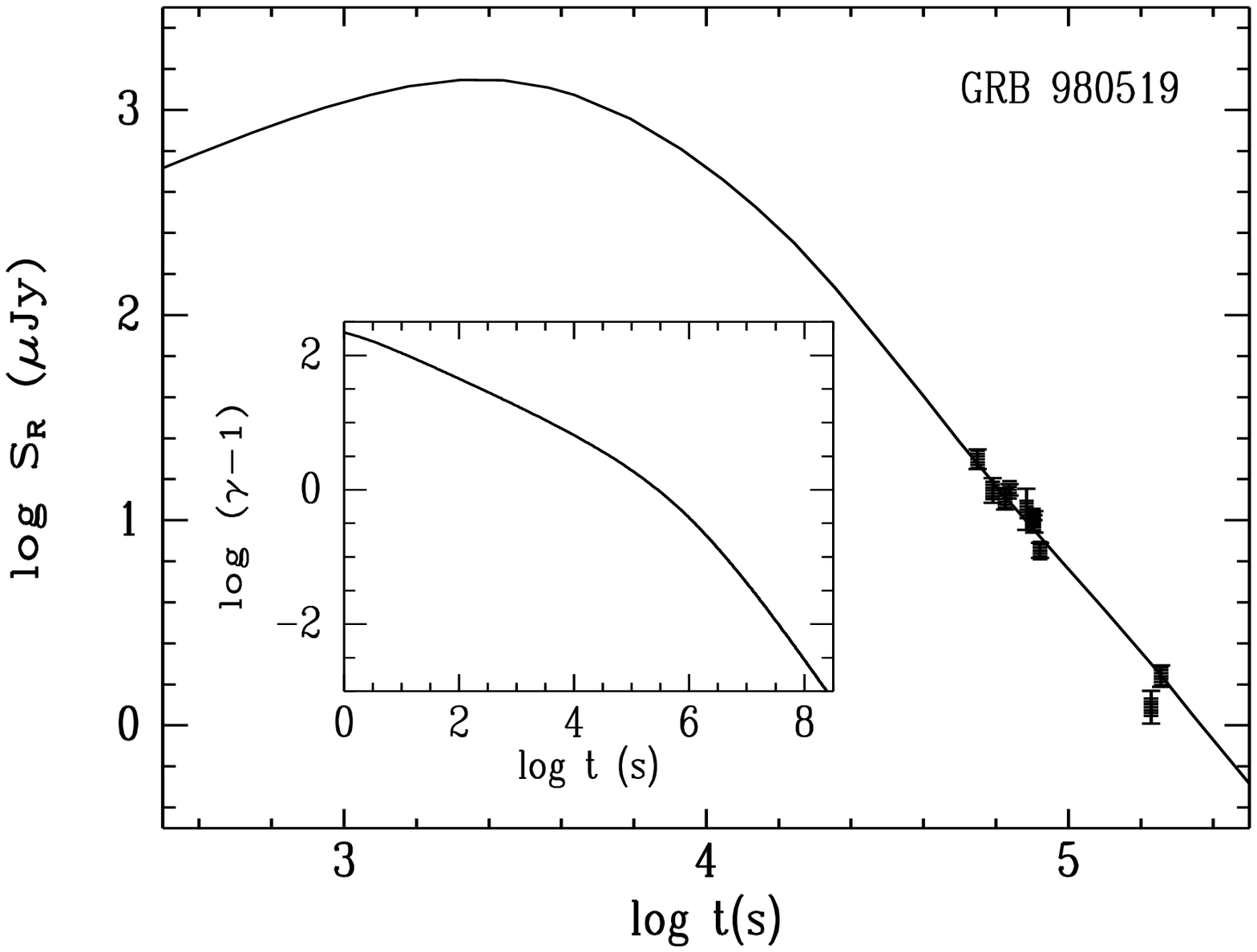,width=2.38in,height=2.0in,angle=-90,
bbllx=180pt, bblly=250pt, bburx=560pt, bbury=710pt}
  \hspace{0.5in}
  \epsfig{figure=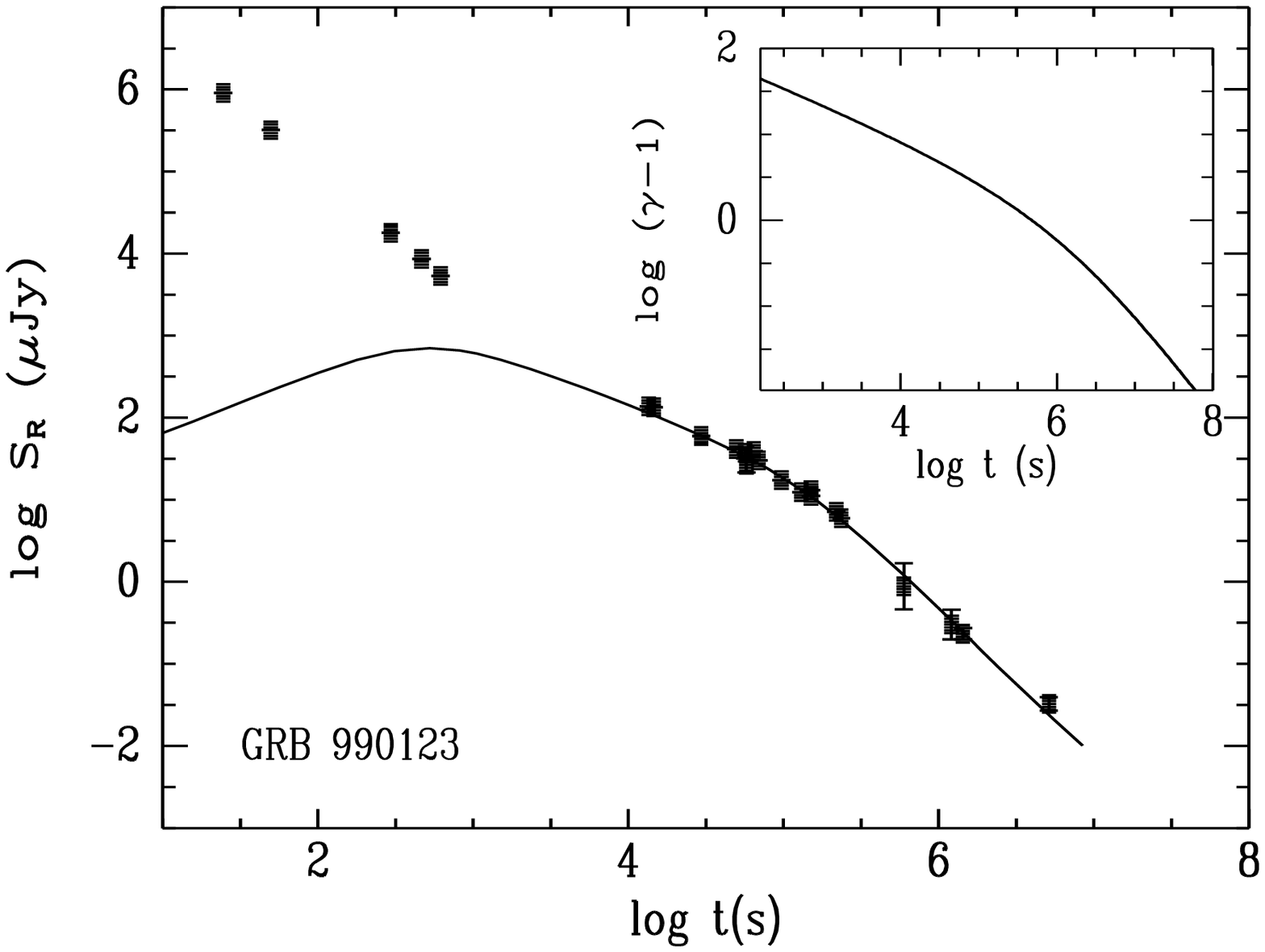,width=2.38in,height=2.0in,angle=-90,
bbllx=180pt, bblly=190pt, bburx=560pt, bbury=650pt}
  }}
  \centerline{ \hbox{ \hbox{} \hspace{0.5in}
  \epsfig{figure=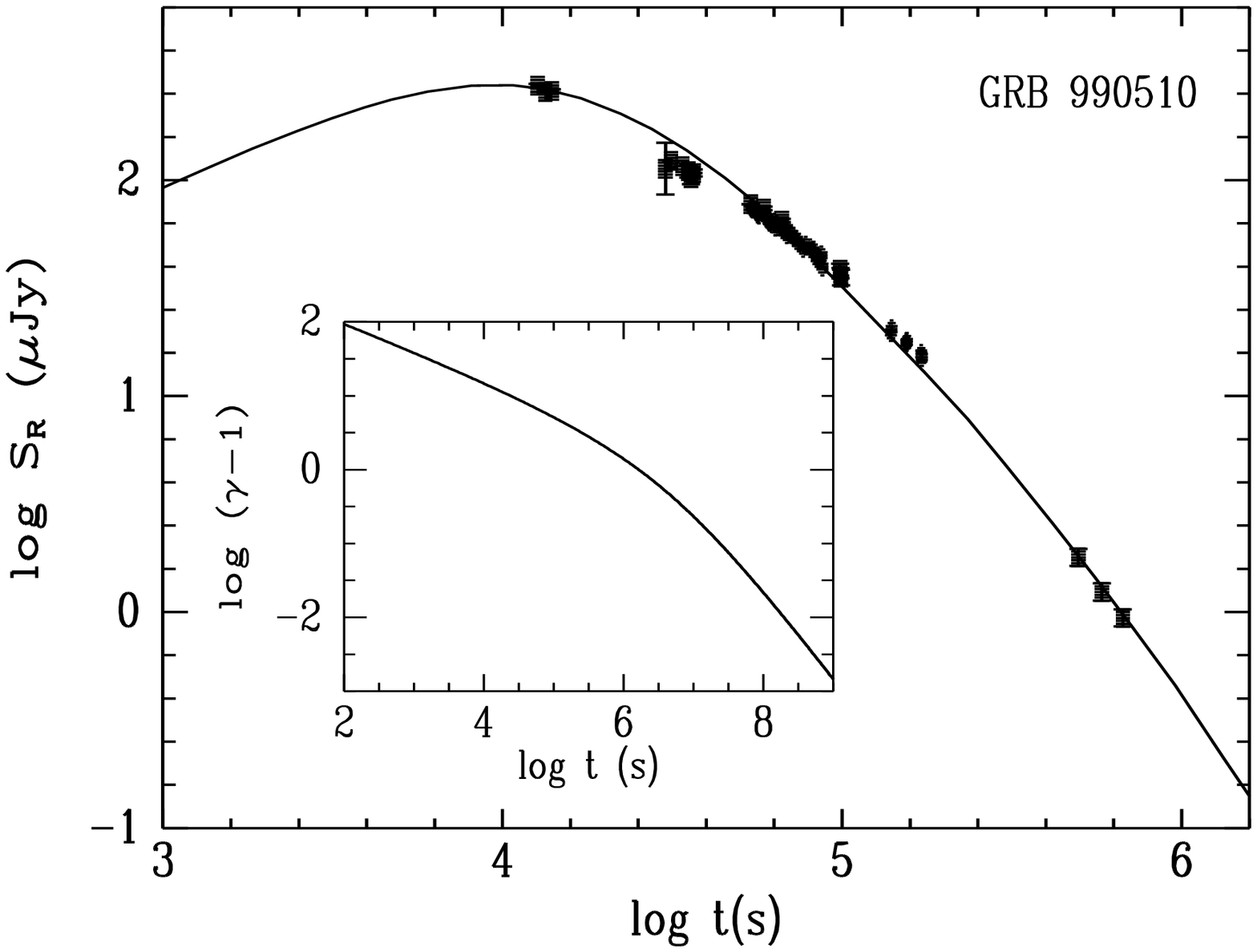,width=2.38in,height=2.0in,angle=-90,
bbllx=135pt, bblly=250pt, bburx=515pt, bbury=710pt}
  \hspace{0.5in}
  \epsfig{figure=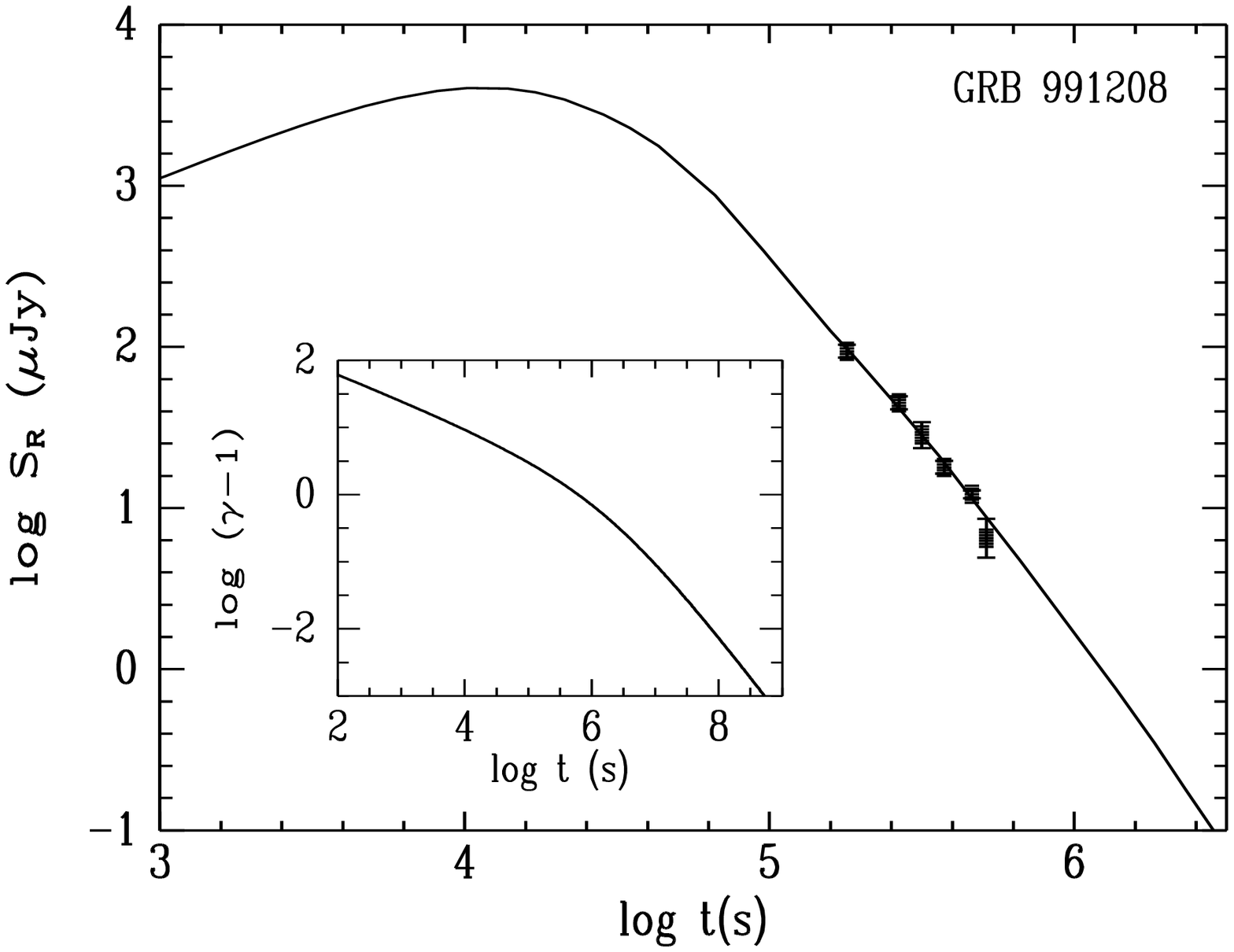,width=2.38in,height=2.0in,angle=-90,
bbllx=135pt, bblly=190pt, bburx=515pt, bbury=650pt}
  }}
\begin{minipage}{12cm}
\caption[]{Optical afterglows from GRB 980519, 990123, 990510, 
991208 and our best fit to them by employing the refined 
jet model. Observed data are collected from the literature. Insets
show the evolution of $\gamma$ in our models (Huang et al. 2000b).}
\label{ssmsbengps}
\end{minipage}
  \end{center}
  \end{figure}
 
\section{Discussion and Conclusions}

To conclude, in this article, we try to answer the following 
question: does the jet effect lead to any breaks in the afterglow 
light curve? We have shown clearly that the theoretically 
predicted light curve steepen due to the edge effect and the 
lateral expansion effect in fact does not exist, consistent 
with recent numerical results by some other researchers
(Moderski et al. 2000).
Our new finding is that a striking break does appear in the 
light curve at the relativistic-Newtonian transition point. 
However, this break is affected by many parameters, such as 
$\xi_{\rm e}, \xi_{\rm B}^2, \theta_0$ and also $n$. Increase 
of any of them to a large enough value will make the break 
disappear. So, although a sharp break in the light curve is 
a good indicator for jet (note that this break is due to 
the relativistic-Newtonian transition, as pointed out above),
the jet effect does not definitely lead to
sharp breaks in the afterglow light curves. This seems to 
make the task of distinguishing jets from isotropic fireballs 
even more difficult.  

Fortunately we have another helpful tool.
Although whether the break appears or not depends on parameters, 
afterglows from jetted GRB remnants are uniformly characterized 
by a quick decay during the non-relativistic phase, with 
$\alpha \geq 2.1$. This is quite different from isotropic 
fireballs. We strongly suggest that 
it can be regarded as a fundamental characteristic of 
jet and may be used to judge the degree of beaming observationally. 
For example, the rapid fading of optical afterglows from 
GRB 970228, 980326, 980519, 990123, 990510 and 991208 has 
been argued as evidence for beaming in these GRBs (Huang et al.
2000b).

Showing only a single flat line (with $\alpha \sim 1.1$) in the 
optical light curve, radiation from GRB 971214 and GRB 980703 is 
not likely be highly collimated. Since they have indicated 
isotropic energies of $\sim 0.17$ M$_{\odot} c^2$ and 
$\sim 0.06$ M$_{\odot} c^2$ respectively (Kulkarni et al. 1998;
Bloom et al. 1998), which
could hardly be met by any kind of compact stellar 
objects, we see that the energy crisis is really a problem. 
 
\vspace{1cm}

\centerline{\Large\bf Reference}

\noindent
Andersen M.I. {\em et al.}, 1999, Sci, 283, 2075

\noindent
Bloom J.S. {\em et al.}, 1998, ApJ, L21

\noindent
Castro-Tirado A. {\em et al.}, 1999, Sci, 283, 2069

\noindent
Chiang J. \& Dermer C.D., 1999, ApJ, 512, 699

\noindent
Costa E. {\em et al.}, 1997, Nat, 387, 783

\noindent
Galama T.J. {\em et al.}, 1999, Nat, 398, 394

\noindent
Groot P.J. {\em et al.}, 1998, ApJ, 523, L121

\noindent
Halpern J.R. {\em et al.}, 1999, ApJ, 517, L105

\noindent
Harrison F.A. {\em et al.}, 1999, ApJ, 523, L121

\noindent
Huang Y.F., Dai Z.G. \& Lu T., 1998a, A\&A, 336, L69

\noindent
Huang Y.F., Dai Z.G. \& Lu T., 1998b, MNRAS, 298, 459 

\noindent
Huang Y.F., Dai Z.G. \& T. Lu, 1999, MNRAS, 309, 513.

\noindent
Huang Y.F. {\em et al.}, 2000a, ApJ, in press
               (astro-ph/9910493)

\noindent
Huang Y.F., Dai Z.G. \& Lu T., 2000b, A\&A, 355, L43
               (astro-ph/0002433)

\noindent
Huang Y.F., Dai Z.G. \& Lu T., 2000c, MNRAS, in press
               (astro-ph/0005549)

\noindent
Huang Y.F., Dai Z.G. \& Lu T., 2000d, Chin. Phys. Lett.,
               in press

\noindent
Kulkarni S.R. {\em et al.}, 1998, Nat, 393, 35

\noindent
Kulkarni S.R. {\em et al.}, 1999, Nat, 398, 389

\noindent
Lamb D.Q., 2000, astro-ph/0005028

\noindent
M\'{e}sz\'{a}ros P. \& Rees M.J., 1992, MNRAS, 257, 29P

\noindent
M\'{e}sz\'{a}ros P. \& Rees M.J., 1999, MNRAS, 306, L39

\noindent
Moderski R., Sikora M. \& Bulik T., 2000, ApJ, 529, 151

\noindent
Owens A. {\em et al.}, 1998, A\&A, 339, L37

\noindent 
Panaitescu A. \& M\'{e}sz\'{a}ros P., 1999, ApJ, 526, 707

\noindent
Piran T., 1999, Phys Rep, 314, 575 

\noindent
Rhoads J., 1999, ApJ, 525, 737

\noindent
Sari R., Narayan R. \& Piran T., 1996, ApJ, 473, 204

\noindent
Sari R., Piran T. \& Halpern J.P., 1999, ApJ, 519, L17

\noindent
Tavani M., 1997, ApJ, 483, L87

\noindent
Vietri M., 1997, ApJ, 488, L105

\noindent
van Paradijs J., Kouveliotou C. \& Wijers R.A.M.J., 
2000, ARA\&A in press

\noindent 
Waxman E., 1997, ApJ, 485, L5

\noindent 
Wei D.M. \& Lu T., 1999a, astro-ph/9908273

\noindent 
Wei D.M. \& Lu T., 1999b, astro-ph/9912063

\noindent
Wijers R., Rees M.J. \& M\'{e}sz\'{a}ros P., 1997, MNRAS, 288, L51

\end{document}